\documentclass{moriond}

\def\be{\begin{equation}}
\def\ee{\end{equation}}
\def\bea{\begin{eqnarray}}
\def\eea{\end{eqnarray}}

\newcommand{\Photo}{}

\begin{document}
DO-TH  16/11
\vspace*{4cm}
\title{$B \to K^* \ell \ell$ Standard Model contributions -- Zooming in on high $q^2$}

\author{G. Hiller}

\address{Institut f\"ur Physik, Technische Universit\"at Dortmund, D-44221 Dortmund, Germany}

\maketitle\abstracts{
To further precision studies with $B \to K^{(*)} \ell \ell$ decays in the 
high-$q^2$ window uncertainties related to  the operator product expansion (OPE) need to be scrutinized.
How well can the OPE describe $B \to K^* (\to K \pi)  \ell \ell$ angular distributions for a given  binning in view of the local charm resonance structure?
We present a data-driven method to access this
quantitatively.  Our analysis suggests that 
the  bins which  are near the kinematic endpoint are best described by the OPE
and should be pursued for precision studies. At the same time
measurements with finer binning help  controlling the uncertainties.}

\section{Introduction}

Rare decays of $B$-mesons into leptons are key modes to test the Standard Model (SM)
and look for New Physics \cite{Blake:2015tda}. In particular $B \to K^* (\to K \pi) \mu \mu$ decays have received high and growing interest due to their  sensitivity to flavor physics in and beyond the SM and the feasibility for precision studies at hadron and $e^+ e^-$-colliders.
Recent  data from the LHC cover several  thousands of events into muons, and include various angular distributions  \cite{Khachatryan:2015isa} \cite{Aaij:2015oid} \cite{LHCb-BR}. There are great prospects
for  Run II and future machines
\cite{Aushev:2010bq}, including various  other final lepton species ($e, \tau, \nu$).

$B \to K^{(*)}  \ell \ell$ decays are described by $1/m_b$-methods, in both the region of  low dilepton mass squared $q^2$, below the $J/\Psi$, and
the high-$q^2$ region above the $\Psi^\prime$.
In the latter is  $q^2 \sim {\cal{O}}(m_b^2)$ and an operator product  expansion (OPE) applies \cite{Grinstein:2004vb}.
Among its benefits is the good convergence, power corrections linear in $1/m_b$ receive additional parametric suppression,
and the resulting universality of short-distance coefficients $ C^{L,R}$ in the  longitudinal $(0)$, parallel $(\parallel)$ and perpendicular $(\perp)$ transversity amplitudes \cite{Bobeth:2010wg}
\begin{equation}
A_j^{L,R} (q^2) \simeq  \mathcal{C}^{L,R}(q^2) f_j(q^2)  + {\cal{O}}(1/m_b)\, , \quad \quad j=0, \parallel, \perp \, .
\label{amplitudes}
\end{equation}
Here,
\begin{equation}
\mathcal{C}^{L,R}(q^2)=\mathcal{C}_9^{\rm eff} (q^2) \mp \mathcal{C}_{10}+ 
      \kappa\frac{2 m_b m_B }{q^2}\, \mathcal{C}_7^{\rm eff}(q^2) \, , 
\end{equation}
where $\mathcal{C}_i$ are the Wilson coefficients of the radiative and semileptonic operators, respectively,
\begin{equation}\label{EffHamiltonian2}
\mathcal{O}_7=\frac{e}{16\pi^2}m_b \bar{s}\sigma^{\mu\nu}P_R b F_{\mu\nu},\quad
 \mathcal{O}_9=\frac{e^2}{16\pi^2}(\bar{s}\gamma^{\mu}P_L b)(\bar\ell\gamma_\mu\ell),\quad
\mathcal{O}_{10}=\frac{e^2}{16\pi^2}(\bar{s}\gamma^{\mu}P_L b)(\bar \ell\gamma_\mu\gamma_5\ell) \, ,
\end{equation}
and the $f_j$ are transversity form factors. The effective coefficients $\mathcal{C}_{7,9}^{\rm eff}$ equal $\mathcal{C}_{7,9}$ up to contributions from 4-quark operators.
Including right-handed currents from BSM physics,  the universality of amplitudes Eq.~(\ref{amplitudes}) breaks down to a partial one between the 
$(0)$ and $(\parallel)$ amplitudes.

\begin{figure}
\centerline{\includegraphics[width=0.4\linewidth]{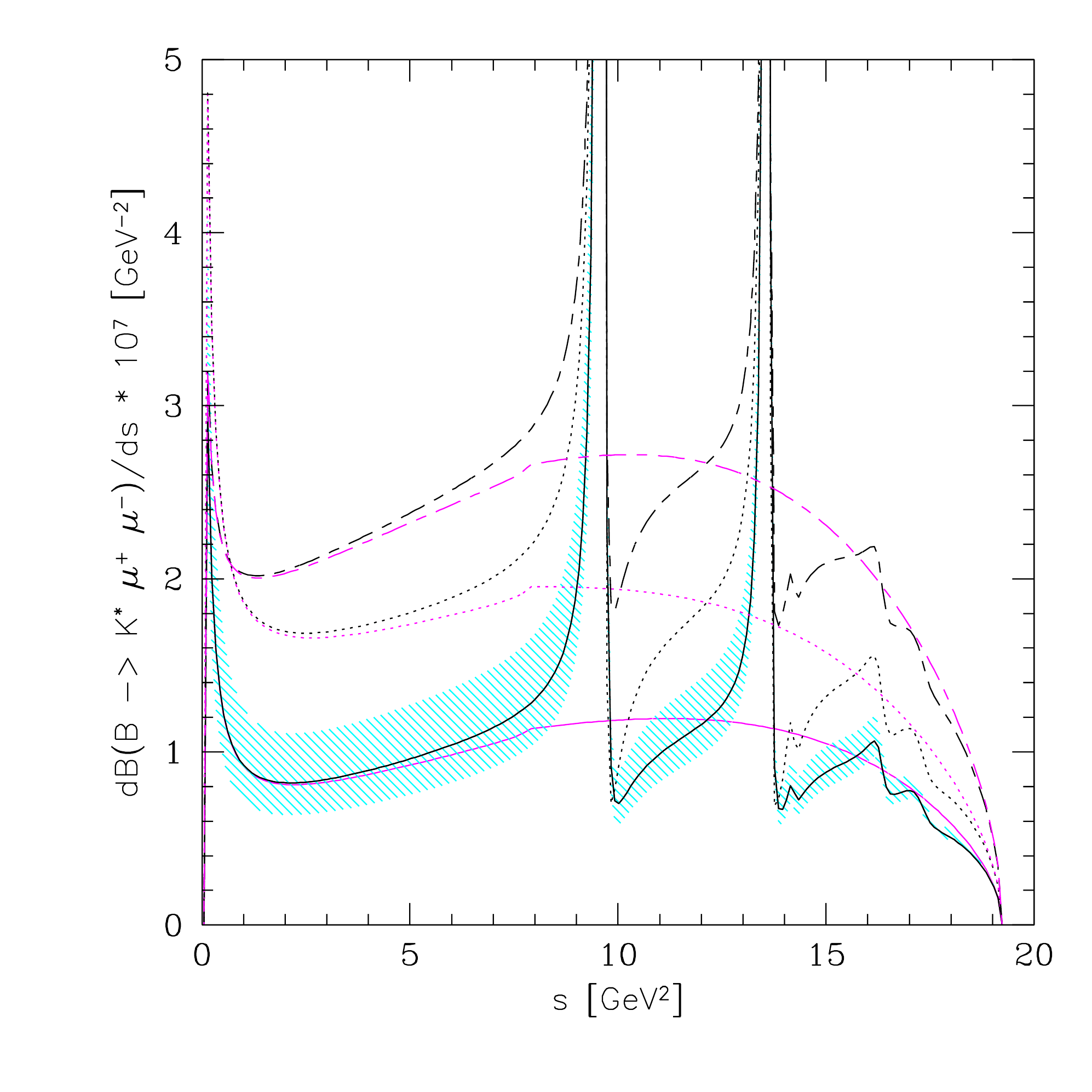}}
\caption[]{The differential branching fraction $d {\cal{B}}/dq^2 (B \to K^* \mu \mu)$ in the SM  (solid curves) and BSM scenarios (dotted and dash-dotted curves). Black curves include resonances based on the KS-approach
\cite{Kruger:1996cv} while the magenta ones are short-distance only. The high-$q^2$ region begins  above the $\psi^\prime$-peak, note the wiggles.
Figure taken from \cite{Ali:1999mm}. }
\label{fig:BKLL}
\end{figure}

As the OPE relies on duality  \cite{Beylich:2011aq}  its perforance depends on bin size and location.
In particular, too small bin intervals will start to resolve local resonance structure induced by  charm contributions $B \to K^* (c \bar c) \to K^* \ell \ell$, shown in Figure \ref{fig:BKLL},
and  limit the accuracy of the OPE predictions at high $q^2$.
As we can't tell this from within the OPE, we employ a local model parametrization  as a test-case
against  the OPE.

\section{The high-$q^2$ region locally}

Charm contributions are electromagnetically induced and modify the operator with vector coupling to leptons, $\mathcal{O}_9$.
The OPE  covers such effects; they are part of the effective coefficient $\mathcal{C}_9^{\rm eff}$ and read, up to terms of order $\alpha_s$ and neglecting
contributions from non-charm penguin operators,
\begin{eqnarray}
\mathcal{C}^{\rm eff}_9(q^2) =\mathcal{C}_9+h(q^2,m_c^2 ) \bigg[\frac{4}{3}\mathcal{C}_1+\mathcal{C}_2+6\mathcal{C}_3+60\mathcal{C}_5\bigg] + \ldots
.\label{C9OPE}
\end{eqnarray}
The loop function $h(q^2,m_c^2)$ is obtained perturbatively from insertions of $\bar s b \bar c c$-type  operators and is smooth in the high-$q^2$ region \cite{Grinstein:2004vb}.

\begin{figure}
 \begin{minipage}{0.33\linewidth}
\centerline{\includegraphics[width=0.9\linewidth]{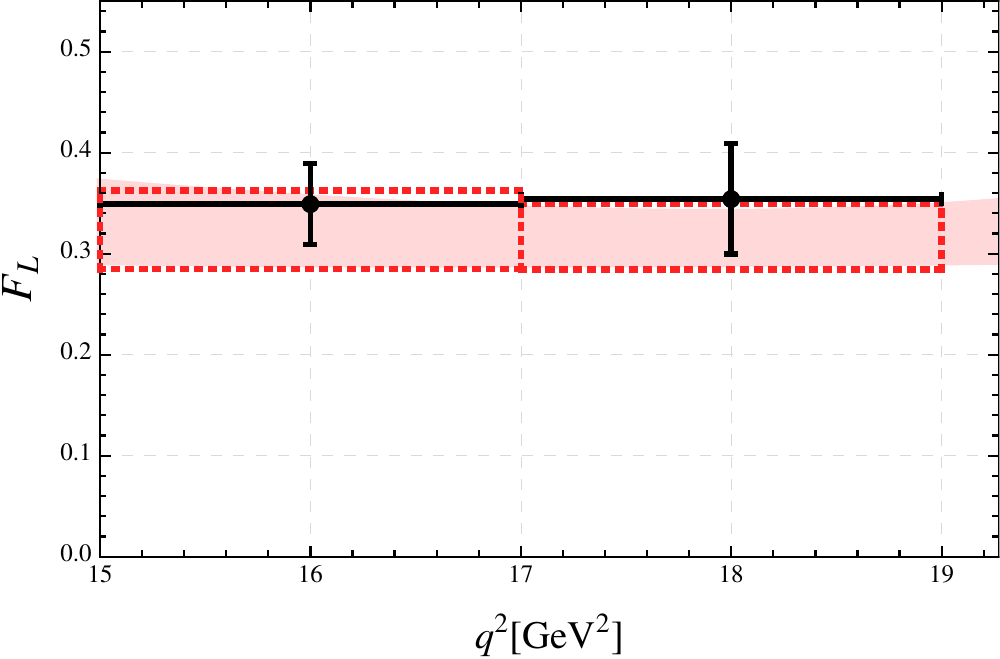}}
\end{minipage}
\hfill
\begin{minipage}{0.32\linewidth}
\centerline{\includegraphics[width=0.9\linewidth]{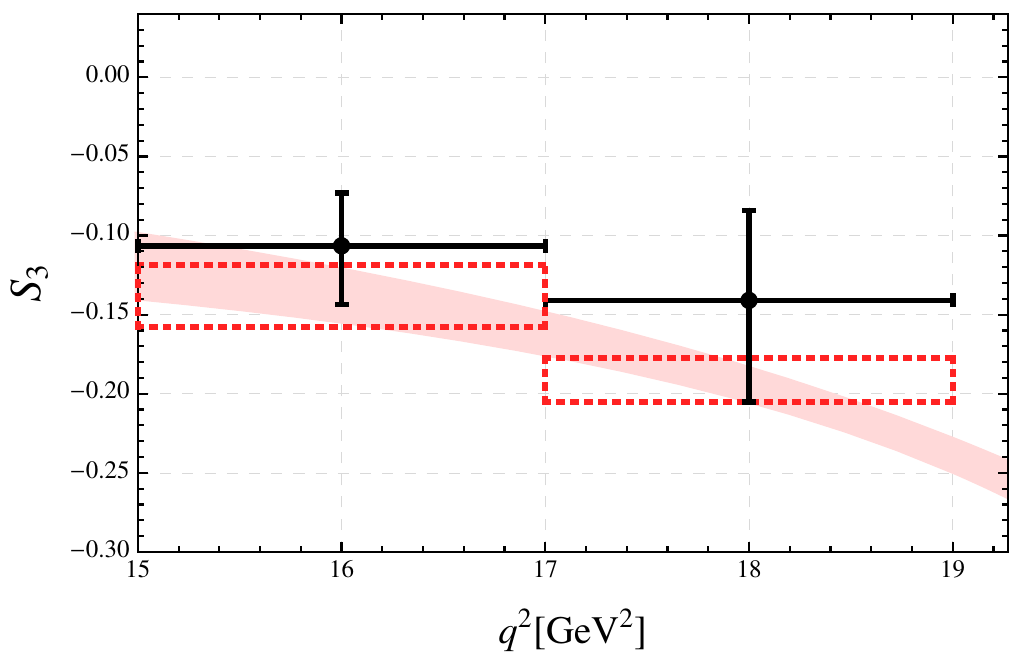}}
\end{minipage}
\hfill
\begin{minipage}{0.32\linewidth}
\centerline{\includegraphics[angle=0,width=0.9\linewidth]{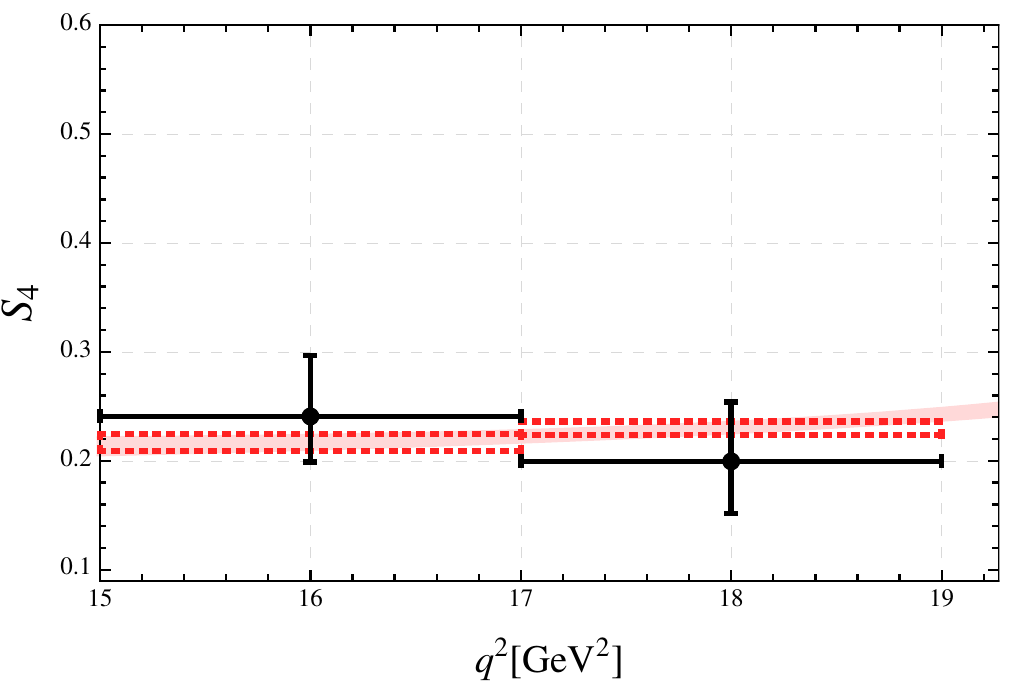}}
\end{minipage}
 \begin{minipage}{0.33\linewidth}
\centerline{\includegraphics[width=0.9\linewidth]{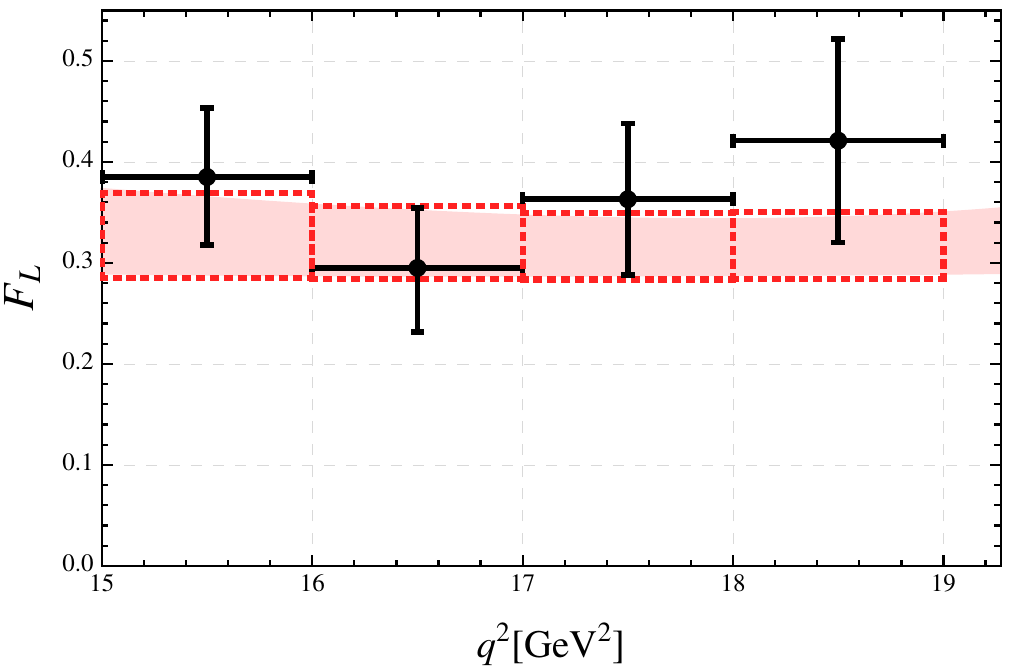}}
\end{minipage}
\hfill
\begin{minipage}{0.32\linewidth}
\centerline{\includegraphics[width=0.9\linewidth]{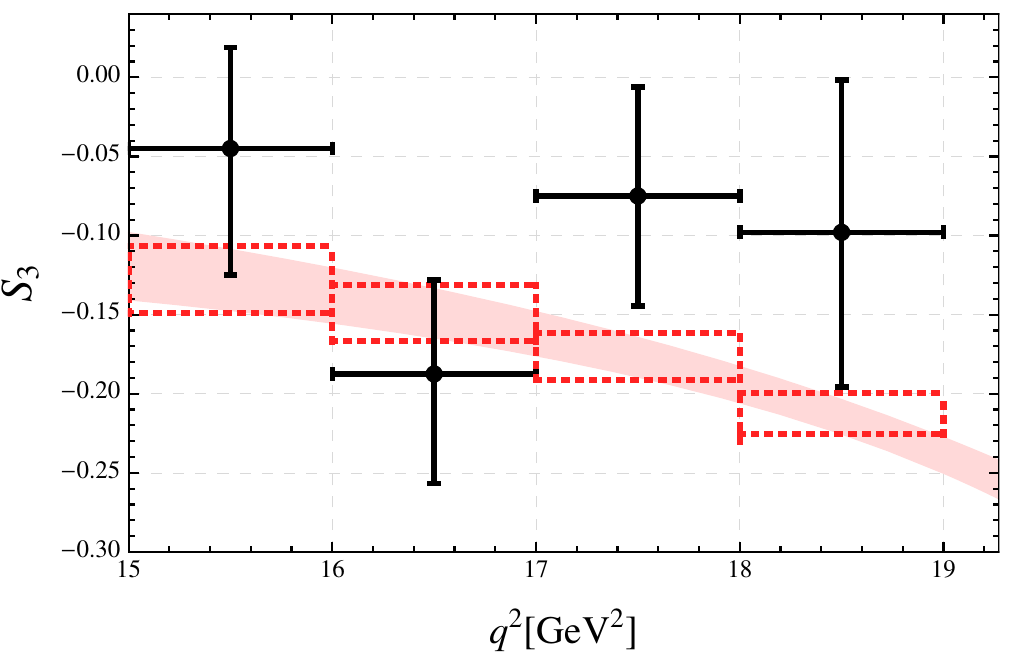}}
\end{minipage}
\hfill
\begin{minipage}{0.32\linewidth}
\centerline{\includegraphics[angle=0,width=0.9\linewidth]{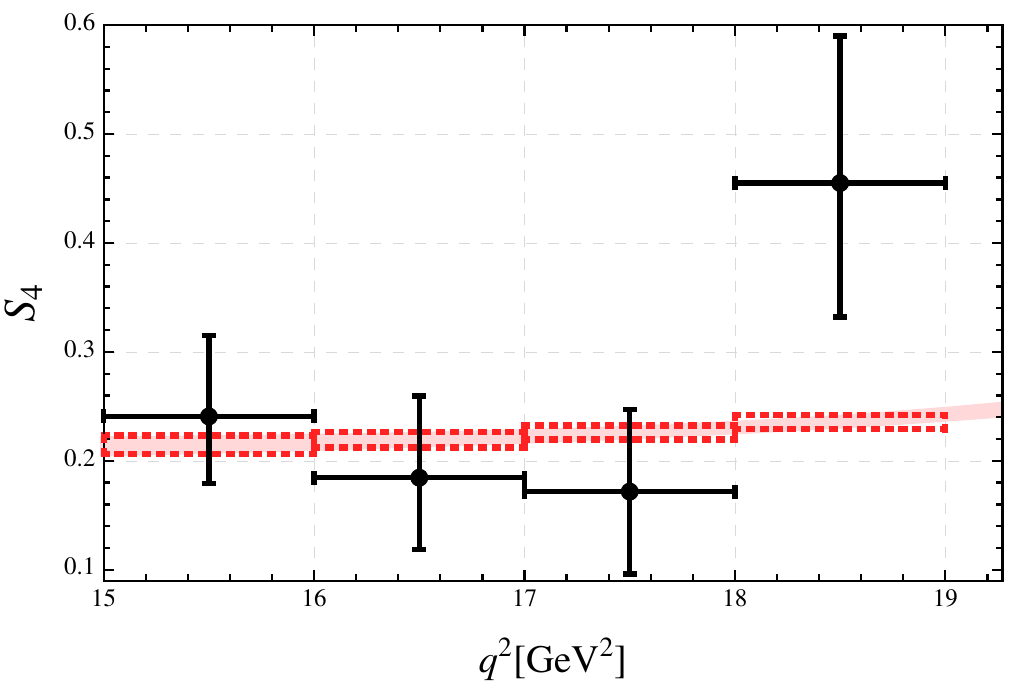}}
\end{minipage}
\caption[]{$F_L, S_{3}$ and $S_4$ in the OPE for 2 $\mbox{GeV}^2$ bins (upper plots) and 1 $\mbox{GeV}^2$ bins (lower plots) 
shown as red boxes versus data (black) from LHCb \cite{Aaij:2015oid}. Systematic and statistical uncertainties are added in quadrature.
The light-shaded red bands illustrate the OPE for infinitesimal binning. Plots taken from \cite{BHN16}.}
 \label{PlotsFLcont}
\end{figure}

In Figure~\ref{PlotsFLcont} the   "short-distance-free" observables $S_3,S_4$ and $F_L$ in  the OPE (red curves and boxes with form factors from~\cite{Horgan:2013hoa}) are compared to data (black, converted to theory conventions),  zooming in from 2 $\mbox{GeV}^2$ bins (upper plots) to  1 $\mbox{GeV}^2$ bins (lower plots).
 Quite generally  one expects from the $R$-ratio \cite{Ablikim:2007gd} an onset of resonance structure with this resolution.
 Indeed the alternating patterns in the 1 $\mbox{GeV}^2$ bins may be hinting at resonances,
however,  due to the limited experimental precision, one cannot draw  firm conclusions presently.
Plots with further observables, $S_5, A_{\rm FB}$ and $d {\cal{B}}/dq^2$, are given in \cite{BHN16}.

 Furthermore, the data have to meet the endpoint relations at $q^2_{\rm max}=(m_B-m_{K^*})^2$, which follow from Lorentz-invariance and hold irrespective of the underlying electroweak model  \cite{Hiller:2013cza} 
\begin{equation} \label{eq:end}
F_L(q^2_{\rm max})=1/3 \, , ~~S_3(q^2_{\rm max})=-1/4 \, , ~~S_4(q^2_{\rm max})=1/4 \, , \, ~~S_{5,6,7,8,9}(q^2_{\rm max})=0  \, .
\end{equation}
 In particular with 1 $\mbox{GeV}^2$ bins  data on $S_{3,4,5}$ are  presently in mild conflict with the endpoint relations.
 Further data with improved precision is required to clarify these points.

\begin{figure}[h]
\includegraphics[width=0.48\textwidth]{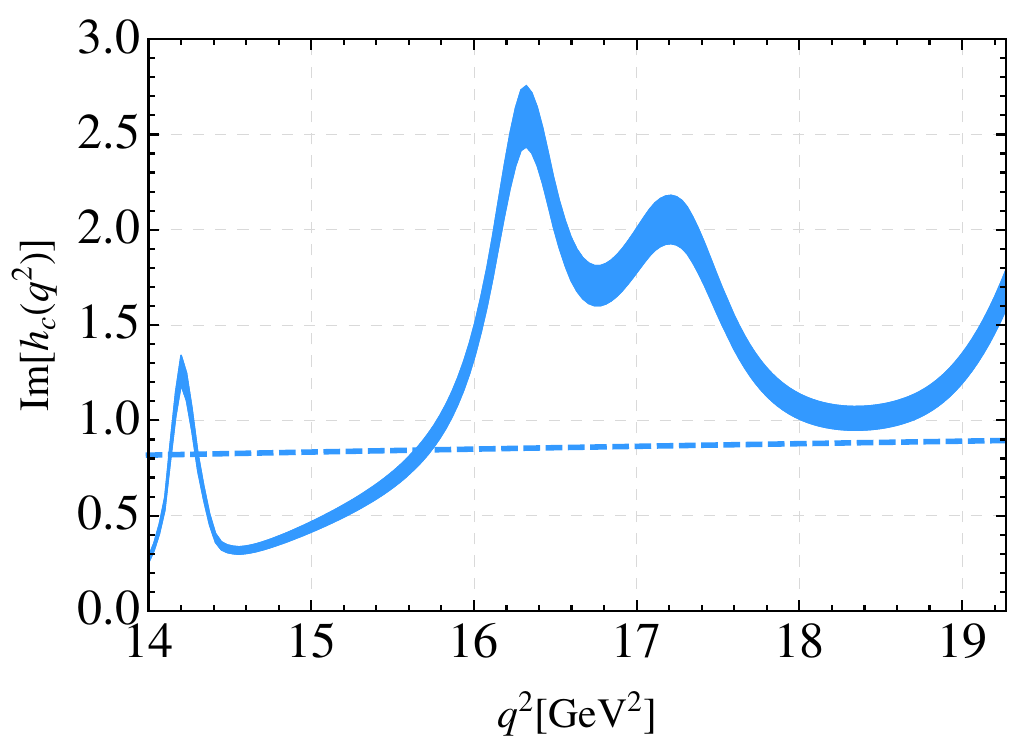}
\includegraphics[width=0.497\textwidth]{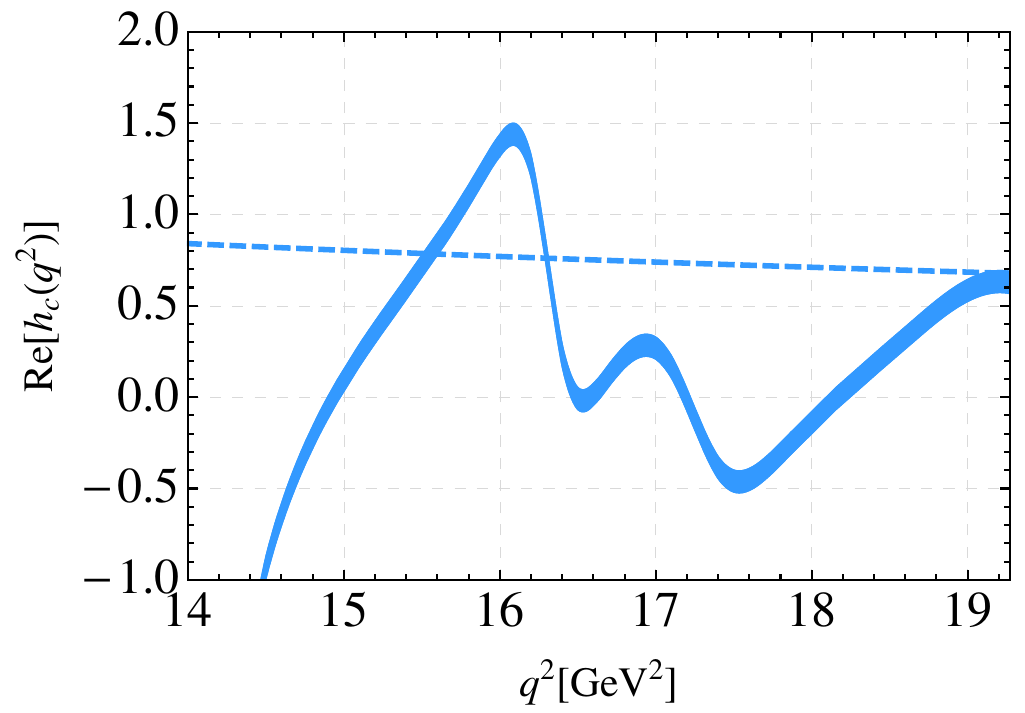}
\caption[]{The charm polarization function $h_c(q^2)$ from  $e^+e^-\to hadrons$ data  (blue $1 \sigma$ band). The  corresponding OPE-contributions,  imaginary and  real part of  $h(q^2,m_c^2)$, are shown by the blue dashed lines. Plots taken from \cite{BHN16}.} 
\label{Fig.1}
\end{figure}

To test the accuracy of the OPE for a given binning we employ a model that is able to locally capture resonance effects which show up  as "wiggles" in the $B \to K^{*} \ell \ell$ distributions.
We follow the  method of Kr\"uger and Sehgal (KS) \cite{Kruger:1996cv}, which uses $e^+ e^-$ -data on the vacuum polarization
to describe the charm-loop function as
\begin{eqnarray}
\mathcal{C}_9^{{\rm eff } }(q^2) \vert^{KS} &=\mathcal{C}_9+ (3 a_2) \, \eta_c \, h_c(q^2) +\ldots,
\label{eq:C9effKS}
\end{eqnarray}
where
\begin{equation}
{\rm Im}   \, h_c(q^2)= \frac{\pi}{3} R_c(q^2) \, , \quad \quad R_c=\frac{\sigma(e^+ e^- \to  c\bar c )}{ \sigma(e^+ e^- \to \mu^+ \mu^-)} \, .
\end{equation}
The real part  of $h_c(q^2) $ is obtained from a dispersion integral.
The  real and imaginary part of $h_c$ extracted from a fit to  BES data~\cite{Ablikim:2007gd} can be seen in Figure \ref{Fig.1}, consistent with
 \cite{Lyon:2014hpa}.
$a_2$ is the same combination of Wilson coefficients that accounts for the perturbative charm-loop
\begin{equation} \label{eq:a2}
a_2=\frac{1}{3}\bigg(\frac{4}{3}\mathcal{C}_1+\mathcal{C}_2+6 \mathcal{C}_3 + 60 \mathcal{C}_5\bigg) \, .
\end{equation}
Numerically, to NNLO accuracy at the $b$-mass scale, $a_2=0.2$.
(In the operator basis used in earlier works $3 a_2$ corresponds to $C^{(0)}$ \cite{Ali:1999mm}.)
The factor $\eta_c$ has been introduced to account for corrections from  beyond naive factorization, $\eta_c=1$.
Such effects are expected quite generally as $B \to K^{(*)} (c \bar c)$ decays are prominent examples of modes with violent breaking of naive factorization \cite{Diehl:2001xe}, yet
a comparison of the factorization formula with measured branching ratios yields
$| \eta_{(J/\psi, \Psi(2S))\,K^\ast}\vert \simeq 0.9-1$.

We go beyond 
the original works \cite{Kruger:1996cv} and generalize Eq.~(\ref{eq:C9effKS}) by introducing transversity-dependent fudge functions $\eta_c(K^*_j,q^2)$ to obtain a model that  can fit  $B \to K^* \ell \ell$ distributions in principle with any precision limited only by input other that of the resonances, such as perturbative one and form factors.
Note that 
symmetries at the endpoint  dictate \cite{Hiller:2013cza}
\begin{equation} \label{eq:EP}
\eta_c(K^*_0, q^2_{\rm max})=\eta_c(K^*_\parallel, q^2_{\rm max}) \, ,
\end{equation}
but other than that the  functions should be constrained experimentally.

To make progress we use constant $\eta_\perp$ and $\eta_\parallel=\eta_0$ to comply with Eq.~(\ref{eq:EP}). We define $\eta_j \equiv  \eta_c(K^*_j, q^2)$. With improved data one can consider different shapes.

\section{Fitting $B \to K^* \mu \mu$ observables}

One benefits from the availablility of $B \to K^* (\to K \pi) \mu \mu$  data \cite{Aaij:2015oid} in different binings 
\begin{equation} \label{eq:bins}
[15-19]  \, \mbox{GeV}^2 \, ,  \quad  [15-17], [17-19]  \,  \mbox{GeV}^2 \, ,\quad    [15-16], \dots ,  [18-19]  \,  \mbox{GeV}^2
\end{equation}
allowing to zoom in with resolution $ \Delta q^2=4,2$ and $1  \mbox{GeV}^2 $, respectively.
As we assume new physics at the electroweak scale and beyond,
{\it a binning-related effect is due to resonances, not New Physics}.

\begin{figure}
\centerline{\includegraphics[width=0.4\linewidth]{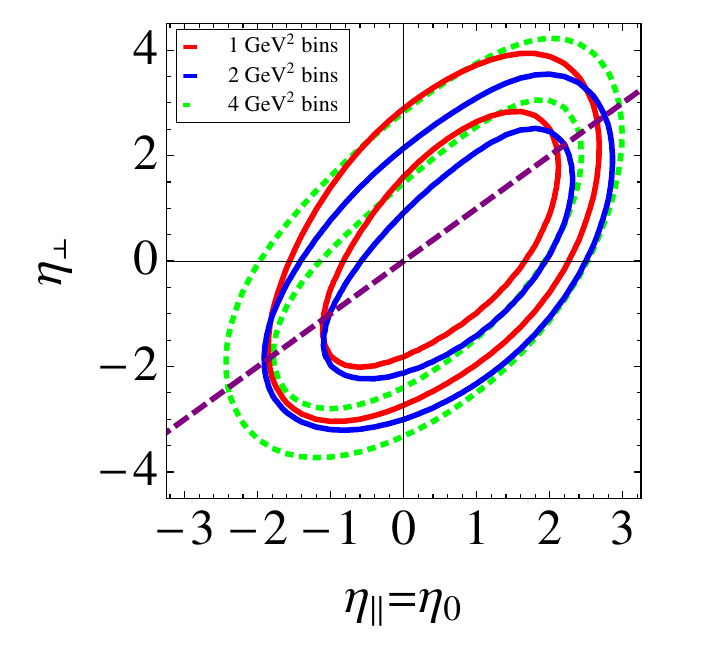}}
\caption[]{$1$ and $2 \, \sigma$ constraints on $\eta_\perp, \eta_\parallel=\eta_0$ from $B \to K^* \mu \mu$ data \cite{Aaij:2015oid} at high $q^2$  for fixed binning 4 $\mbox{GeV}^2$(green, dotted), 2 $\mbox{GeV}^2$  (blue) and 
1 $\mbox{GeV}^2$ (red) as in Eq.~(\ref{eq:bins}). The dashed magenta straight line denotes the universality-limit
$\eta_\perp=\eta_0=\eta_\parallel$. Plot taken from \cite{BHN16}.}
\label{fig:eta}
\end{figure}

\begin{figure}
\begin{minipage}{0.33\linewidth}
\centerline{\includegraphics[width=0.9\linewidth]{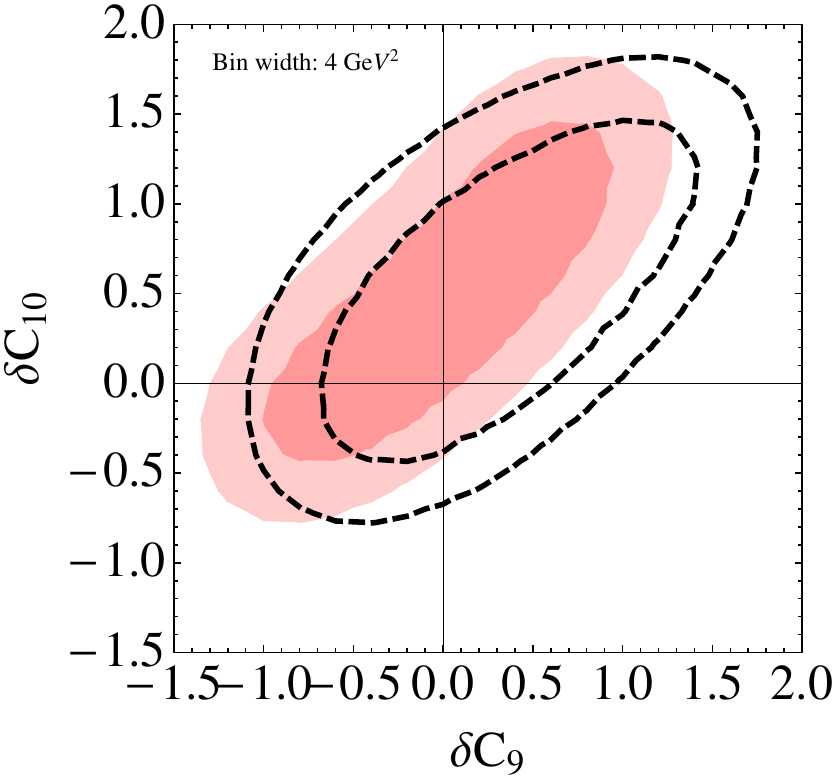}}
\end{minipage}
\hfill
\begin{minipage}{0.32\linewidth}
\centerline{\includegraphics[width=0.9\linewidth]{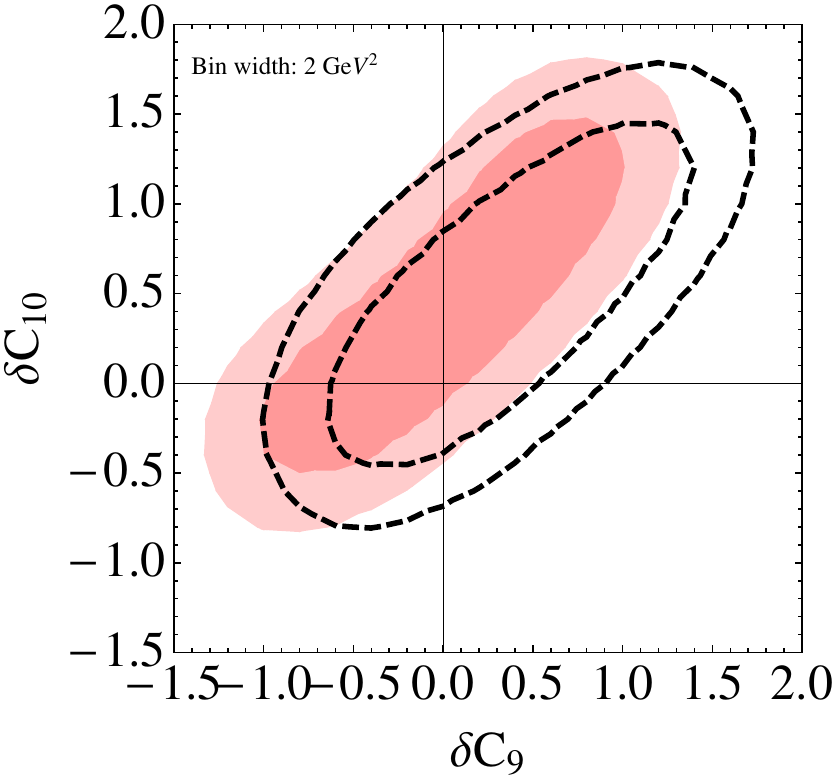}}
\end{minipage}
\hfill
\begin{minipage}{0.32\linewidth}
\centerline{\includegraphics[angle=0,width=0.9\linewidth]{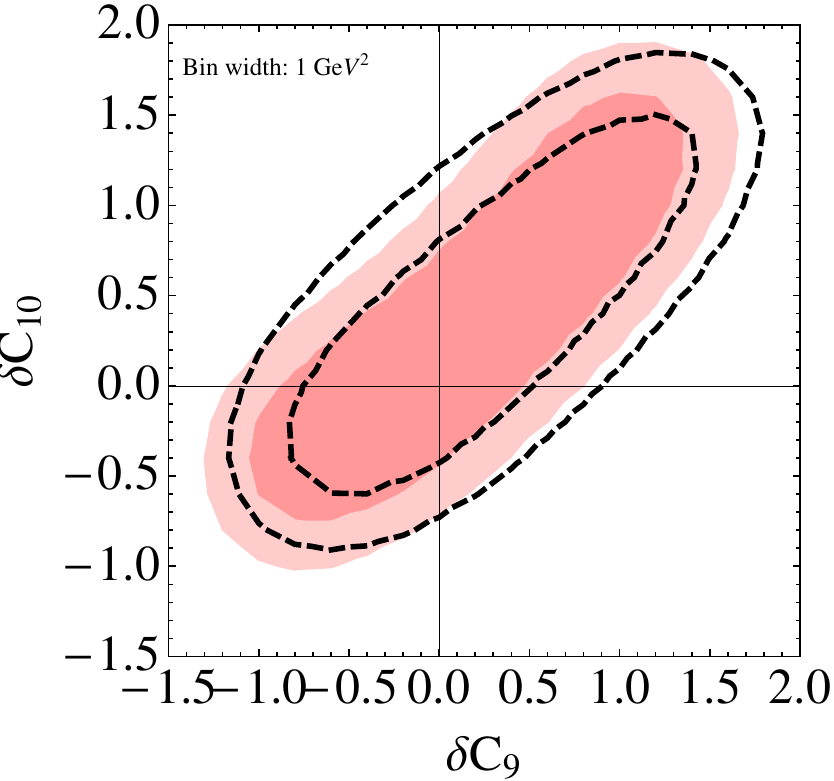}}
\end{minipage}
\caption[]{$1$ and $2 \, \sigma$ constraints on $\delta \mathcal{C}_9, \delta \mathcal{C}_{10}$ from $B \to K^* \mu \mu$ data \cite{Aaij:2015oid}  at high $q^2$  for fixed binning 4 $\mbox{GeV}^2$(left), 2 $\mbox{GeV}^2$  (center) and 
1 $\mbox{GeV}^2$ (right) as in Eq.~(\ref{eq:bins}). Red shaded areas  (black contours) denote the allowed  regions in the OPE (in the KS-approach with $\eta_\perp, \eta_\parallel=\eta_0$ simultaneously fitted). Plots taken from \cite{BHN16}.}
\label{fig:C9C10}
\end{figure}

In addition to $S_3,S_4$ and $F_L$, in which universal effects drop out, we consider 
the angular observables $J_{5, ...,9}$ and the differential branching fraction, $d {\cal{B}}/dq^2$. As these are short-distance dependent, we are forced to perform a joined extraction of
$\eta$'s and $\mathcal{C}$'s.
We simultaneously fit to $\eta_\perp, \eta_\parallel$ and New Physics contributions $\delta \mathcal{C}_9, \delta \mathcal{C}_{10}$ for each $q^2$-resolution
$\Delta q^2=4,2$ and $1 \, \mbox{GeV}^2$.
Results are shown in Figure \ref{fig:eta} for the $\eta$'s and in Figure  \ref{fig:C9C10} for the Wilson coefficients.
As presently there are no data on the branching ratio for  $1  \mbox{GeV}^2 $ bins available, we use in the latter analysis
the $2  \mbox{GeV}^2 $ finding \cite{LHCb-BR}. This is not ideal as it certainly blurs the zooming effect,  however we chose  to keep  the branching ratio as an important constraint in the fit.
Form factors are taken from lattice QCD \cite{Horgan:2013hoa}.

{}From  Figure \ref{fig:eta}  we learn that naive factorization, $\eta_j=1$, is allowed, but also solutions away from universality, $\eta_\perp=\eta_0=\eta_\parallel$, shown by the dashed line.
The constraints from the largest $q^2$-resolution (green, dotted)  are the weakest. The fits are presently consistent with no wiggles, $\eta_j=0$.
However, modulo experimental uncertainties, binning-induced differences hint at the presence of such structure.

Fits to the BSM-coefficients $\delta \mathcal{C}_9, \delta \mathcal{C}_{10}$, see Figure  \ref{fig:C9C10}, give very similar results
for the OPE and the local model,  for each $\Delta q^2$-resolution.
This implies that at  current level of precision, the OPE describes the data sufficiently well
as we are not yet fully sensitive to charm resonances.
The plots also show that the SM is allowed but also sizable BSM effects, in agreement with the plain low recoil analysis of \cite{Descotes-Genon:2015uva}.
An analogous fit to $\eta_\perp, \eta_\parallel$ and the Wilson coefficients of the chirality-flipped operators, $\mathcal{C}^\prime_{9,10}$, gives qualitatively very similar results \cite{BHN16}.

\section{Binning performance}

To estimate the uncertainties of the  OPE prediction, for a given binning, we use the ratios
\begin{equation} \label{eq:ek}
\epsilon_i = \frac{ \int_{bin} \rho_i^{KS} (q^2) dq^2}{ \int_{bin} \rho_i^{OPE} (q^2)  dq^2} \, , ~ i=1,2 \, ,  \quad 
\epsilon_{12} = \frac{ \int_{bin} \rho_2^{KS}  (q^2) dq^2}{ \int_{bin} \rho_2^{OPE} (q^2) dq^2} \cdot    \frac{ \int_{bin} \rho_1^{OPE}  (q^2)dq^2}{ \int_{bin} \rho_1^{KS} (q^2) dq^2}
 \, ,
\end{equation}
where
\begin{eqnarray}
\rho_1(q^2)& \equiv \frac{1}{2}(\vert C^R (q^2)\vert^2+ \vert C^L (q^2)\vert^2)=\bigg\vert\mathcal{C}_9^{\rm eff} (q^2)+\kappa\frac{2 m_b m_B}{q^2}\mathcal{C}_7^{\rm eff} (q^2) \bigg\vert^2+\vert\mathcal{C}_{10}\vert^2 \, , \\
\rho_2(q^2)& \equiv \frac{1}{4}(\vert C^R(q^2) \vert^2- \vert C^L (q^2) \vert^2)={\rm Re}\bigg[\bigg(\mathcal{C}_9^{\rm eff} (q^2)+\kappa\frac{2 m_b m_B}{q^2}\mathcal{C}_7^{\rm eff}(q^2)\bigg)\mathcal{C}^\ast_{10}\bigg] \, ,
\end{eqnarray}
are the short-distance factors at high $q^2$. They have to be evaluated with the respective effective coefficient $C_9^{\rm eff}(q^2)$, Eq.~(\ref{eq:C9effKS})
for the KS-model and Eq.~(\ref{C9OPE}) for the OPE. 
The  closer $\epsilon_k$ to one, the better the performance of the OPE.
We calculate the $\epsilon_k$ model-independently,  {\it  i.e.,} within the global fit, within the  $1 \sigma$ ranges of  Figure \ref{fig:eta} and corresponding $\mathcal{C}_{9,10}$ values.
The outcome is shown  in Table \ref{tab:error}. 
As expected, a larger bin interval and one closer to the kinematic endpoint is best.
The bins with the best  performance are presently  $[17-19]$ and  $[18-19] \, \mbox{GeV}^2$, both near the endpoint, followed by the full one
$[15-19] \, \mbox{GeV}^2$.
The deviations $|\epsilon_k-1|$  from the OPE also include uncertainties within the local charm model.
Therefore, improved understanding of the resonance parameters can reduce the uncertainty on the OPE's performance.
Such information requires measurements in bins that do resolve the wiggles.

\begin{table}
 \caption{ Ranges of   $\epsilon_k$ defined in Eq.~(\ref{eq:ek}) for different $q^2$-bins in $\mbox{GeV}^2$ and $1 \sigma$ ranges of  parameters $\eta_{\perp.\parallel}$, $\mathcal{C}_{9,10}$. }
 \begin{center}
\begin{tabular}{c||c||c|c||c|c|c|c}
 &  $15-19$ & $15-17 $ & $17-19$ &  $15-16$ & $16-17$& $17-18$ & $18-19$  \\ \hline
$\epsilon_1$  &(0.85,1.16) & (0.81,1.30&(0.87,1.03) &(0.76,1.20) &(0.84,1.38) &(0.84,1.03) &(0.86,1.05)\\
$\epsilon_2$  &(0.82,1.0) &(0.74,1.13) &(0.85,0.91) &(0.71,1.17) & (0.78,1.08)&(0.76,0.95) &(0.84,0.97)\\
$\epsilon_{12}$  &(0.86,1.05) &(0.87,1.05) & (0.84,1.05)&(0.95,1.06) &(0.78,1.05) &(0.75,1.05) & (0.93,1.05)\\
\end{tabular}
  \label{tab:error}
 \end{center}
\end{table}

\section{Conclusions}

The high-$q^2$ region in semileptonic rare $|\Delta b| =|\Delta s|=1$ decays is inhabited by wider charm resonances.
Using a local model against the OPE 
provides a data-driven method to
test the  binning and limitations of the OPE.  The $\epsilon_k$ ratios defined in Eq.~(\ref{eq:ek}) 
 provide data-extracted upper limits on  the OPE's  binning-related uncertainty, and are useful to identify the most suitable binning.
 The two bins near the kinematic endpoint perform best,  see Table \ref{tab:error}.
Besides precision studies in these bins  measurements in finer $q^2$-bins are desirable to improve the local description of resonances.
$B \to K^* \mu \mu$ observables at low recoil are presently consistent with the SM,
however, large BSM effects are also allowed.
We   look forward to future data.

\section*{Acknowledgments}

GH is happy to thank her collaborators \cite{BHN16} and the organizers of this inspiring conference for providing an opportunity to speak.
The work reported here has been supported in part by the Bundesministerium f\"ur Bildung und Forschung (BMBF).

\end{document}